\documentclass[a4paper,fleqn]{article}
\usepackage{arxiv}

\usepackage{amssymb}
\usepackage{amsmath}
\usepackage{mathtools}
\usepackage{amsfonts}
\usepackage{upgreek}

\usepackage[numbers]{natbib}
\usepackage{etoolbox}
\usepackage{verbatim} 
\usepackage{svg}
\usepackage{graphicx}
\usepackage{tkz-fct}
\usepackage{multirow}
\usepackage{listings}
\definecolor{listinggray}{gray}{0.9}
\definecolor{Darkgreen}{rgb}{0,0.4,0}
\definecolor{lbcolor}{rgb}{0.9,0.9,0.9}

\usepackage{siunitx}
\usepackage{subcaption}
\usepackage{xcolor}
\usepackage{booktabs,tabularx,longtable}
\usepackage[para,online,flushleft]{threeparttable}

\usepackage{array}
\usepackage{booktabs}
\usepackage{rotating}
\usepackage{bm}

\newcommand\cmc[1]{\omit\hfil\(#1\)\hfil\ignorespaces}
\usepackage{cancel}
\usepackage{blindtext}
\usepackage{scalerel}
\usepackage{hyperref}
\hypersetup{
  colorlinks   = true, 
  urlcolor     = red, 
  linkcolor    = blue, 
  citecolor    = red 
}

\DeclareMathOperator{\curl}{curl}
\DeclareMathOperator{\divL}{div}
\DeclareMathOperator{\grad}{grad}
\DeclareMathOperator{\ellipticK}{K}
\DeclareMathOperator{\ellipticE}{E}

\newcommand{\RF}{{\mathrm{F}}}
\newcommand{\Rd}{{\mathrm{d}}}
\renewcommand{\Re}{{\mathrm{e}}}
\newcommand{\Rf}{{\mathrm{f}}}
\newcommand{\CB}{{\mathcal{B}}}
\newcommand{\CK}{{\mathcal{K}}}
\newcommand{\CP}{{\mathcal{P}}}
\newcommand{\CS}{{\mathcal{S}}}
\newcommand{\SR}{\mathbb{R}}
\newcommand{\OM}{{\Omega}}
\newcommand{\Bom}{{\boldsymbol{\omega}}}

\newcommand{\BB}{{\boldsymbol{B}}}
\newcommand{\BH}{{\boldsymbol{H}}}
\newcommand{\BM}{{\boldsymbol{M}}}

\newcommand{\Be}{{\boldsymbol{e}}}
\newcommand{\Bf}{{\boldsymbol{f}}}
\newcommand{\Bh}{{\boldsymbol{h}}}
\newcommand{\Bm}{{\boldsymbol{m}}}
\newcommand{\Bn}{{\boldsymbol{n}}}
\newcommand{\Bx}{{\boldsymbol{x}}}
\newcommand{\Bzero}{{\boldsymbol{0}}}
\newcommand{\PI}{{\Pi}}
\newcommand{\da}{\,\Rd a}
\newcommand{\dV}{\,\Rd V}
\newcommand{\dv}{\,\Rd v}
\newcommand{\dx}{\,\Rd x}
\newcommand{\dy}{\,\Rd y}
\newcommand{\de}{{\delta}}
\newcommand{\pt}{{\partial}}
\newcommand{\fracpt}[2]{\frac{\pt #1}{\pt #2}}
\newcommand{\hypergeom}[2]{\,_{#1}\RF_{#2}}
\newcommand{\Cpp}{\mbox{C++}\xspace}
\newcommand{\jumpl}{{[\kern-.15em[}}
\newcommand{\jumpr}{{]\kern-.15em]}}
\title{Thin cylindrical magnetic nanodots revisited: variational formulation, accurate solution and phase diagram}

\author{
\href{https://orcid.org/0000-0003-1365-3265}{\includegraphics[scale=0.06]{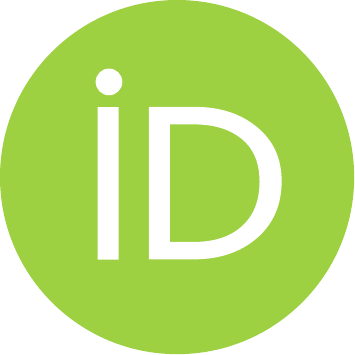}\hspace{1mm}Alexander~ M\"uller}\thanks{Corresponding author} \\
	Institute for Structural Mechanics\\ University of Stuttgart\\
	70550 Stuttgart, Pfaffenwaldring 7 \\
	\texttt{mueller@ibb.uni-stuttgart.de} \\
	\And
 \href{https://orcid.org/0000-0003-1538-4281}{\includegraphics[scale=0.06]{orcid.pdf}\hspace{1mm}Manfred~Bischoff} \\
	Institute for Structural Mechanics\\ University of Stuttgart\\
	70550 Stuttgart, Pfaffenwaldring 7 \\
	\texttt{bischoff@ibb.uni-stuttgart.de} 
 \And 
 \href{https://orcid.org/0000-0002-5838-5201}{\includegraphics[scale=0.06]{orcid.pdf}\hspace{1mm}Marc-Andr\'e~Keip} \\
	Institute of Applied Mechanics\\ University of Stuttgart\\
	70550 Stuttgart, Pfaffenwaldring 7 \\
	\texttt{marc-andre.keip@mechbau.uni-stuttgart.de} 
}
\usepackage[capitalise]{cleveref}
\renewcommand{\shorttitle}{Thin cylindrical magnetic nanodots revisited}
\begin{document}
\maketitle


\begin{abstract}
We investigate the variational formulation and corresponding minimizing energies for the detection of energetically favorable magnetization states of thin cylindrical magnetic nanodots. Opposed to frequently used heuristic procedures found in the literature, we revisit the underlying governing equations and construct a rigorous variational approach that takes both exchange and demagnetization energy into account. Based on a combination of Ritz's method and a Fourier series expansion of the solution field, we are able to pinpoint the precision of solutions, which are given by vortex modes or single-domain states, down to an arbitrary degree of precision. Furthermore, our model allows to derive an expression for the demagnetization energy in closed form for the in-plane single-domain state, which we compare to results from the literature. A key outcome of the present investigation is an accurate phase diagram, which we obtain by comparing the vortex mode's energy minimizers with those of the single-domain states. This phase diagram is validated with data of two- and three-dimensional models from literature. By means of the phase diagram, we particularly find the critical radius at which the vortex mode becomes unfavorable with machine precision. All relevant data and codes related to the present contribution are available at~\cite{darus-3103_2023}.
\end{abstract}

\keywords{
	micromagnetics \and variational formulation \and Ritz method \and vortex mode \and phase diagram
}

\section{Introduction}
\label{sec:intro}
Magnetic materials are usually characterized by the presence of magnetic domains, that is, by regions with equally oriented magnetization. In thin magnetic specimens, the formation of domains is often accompanied by the creation of magnetic vortices~\cite{wachowiakDirectObservationInternal2002,hehnNanoscaleMagneticDomains1996,daoMicromagneticsSimulationDeepsubmicron2001,schneiderLorentzMicroscopyCircular2000}. Such a vortex is characterized by a swirl of magnetization that by default leads to an out-of-plane tilt of magnetization in its core~\cite{anirban2021stable}. As a result of its circular shape, a favorable specimen for the generation of a magnetic vortex is given by a cylindrical magnetic nanodot. Due to the great stability of magnetic vortices in magnetic nanodots, they are potential candidates for magnetic storage devices~\cite{cowburnMagneticNanodotsDevice2002,cowburnChangeDirection2007}. Furthermore, magnetic nanodots are being explored for therapeutical and diagnostical processes in biomedical engineering~\cite{manzin2021micromagnetic}. We refer to~\cite{usovMagnetizationCurlingFine1993,aharoniUpperBoundSingle1990,guslienkoMagnetizationReversalDue2001,leeUniversalCriterionPhase2008,budaMicromagneticSimulationsMagnetisation2002,pigeauMagneticVortexDynamics,gouvaVorticesClassicalTwodimensional1989,brown1963micromagnetics,aharoniSingledomainFerromagneticCylinder1989} for a variety of studies on the statics and dynamics of magnetization in nanodots and to~\cite{guslienkoMagneticVortexState2008,hubertMagneticDomainsAnalysis1998} for a general overview of the topic.

In the present contribution, we are concerned with the analysis and the prediction of magnetization states of thin cylindrical nanodots. In this context, we are motivated by experiments~\cite{cowburnSingleDomainCircularNanomagnets1999,rossMicromagneticBehaviorElectrodeposited2002}
and theoretical studies~\cite{usovMagnetizationCurlingFine1993,guslienkoVortexStateStability2004}.
On the theoretical side, we would like to highlight in particular the contribution of~\citet{usovMagnetizationCurlingFine1993},
which has been serving as a foundation for numerous approaches to study the stability as well as the associated critical radius of the vortex configuration in the static setting. Here, the critical radius is defined as the one at which the vortex mode becomes unstable resulting in constant magnetization throughout the specimen, that is, in a single-domain state. Theoretical upper and lower bounds of the critical radius have been derived in the contributions~\cite{aharoniUpperBoundSingle1990,metlovStabilityMagneticVortex2002,guslienkoVortexStateStability2004}, which are all based on the aforementioned approach~\cite{usovMagnetizationCurlingFine1993}.%
\footnote{%
	Deviating
	from~\cite{usovMagnetizationCurlingFine1993},
	the approaches documented
	in~\cite{aharoniMagnetostaticsCurlingFinite1990,ishiiMagnetizationCurlingFinite1989}
	are based on an ansatz of the magnetization that is given as a function of the thickness coordinate rather than the radial coordinate. Associated approaches are related to a form of curling that is of no interest in the present contribution.
}

In the present work we derive a variational principle based on the stray- and exchange-energy density that provides the full set of micromagnetic governing equations as Euler-Lagrange equations. We exploit this variational principle to investigate energy minimizers with an appropriate Ritz ansatz for the vortex mode. These minimizers are compared with the minimizers obtained from a homogeneous in-plane magnetization. From the latter, we then derive the magnetostatic energy in closed form for varying radii and thicknesses of a nanodot. As a key outcome of the analysis, we eventually compute accurate solutions for the critical radius of a stable vortex state.

Please note that in contrast to existing results found in literature, the results presented herein are either exact or accurate up to machine precision. In this way they perfectly suit as benchmarks for the validation of numerical simulation schemes for magnetic nanodots and related problems of magnetostatics.


\section{Theory}
\subsection{Governing equations and constitutive law}
In what follows, we consider a magnetic body in a configuration $\CB \subset \SR^d$ with boundary $\partial \CB$, where \(\SR^d\) is the \(d\)-dimensional Euclidean space. The body is embedded in free space $\OM \subset \SR^d$ that is assumed to include the space occupied by the magnetic body such that $\CB \subset \OM$. All space is spatially parameterized in the coordinates $\Bx \in \OM$. Then, in a purely magnetostatic setting, the Maxwell equations read
\begin{equation}
	\divL \BB = 0
	\quad \mbox{and} \quad
	\curl \BH = \Bzero
	\quad \mbox{in} \ \OM ,
	\label{eq:maxwellmagnet}
\end{equation}
where \(\BB \) denotes the magnetic induction and \(\BH \) denotes the magnetic field. The Maxwell equations are coupled through the micromagnetic constitutive equation
\begin{equation}
	\BB = \mu_0 ( \BH + M_s \BM ),
	\label{eq:constitutive}
\end{equation}
where \(\mu_0\) and \(M_s\) denote the magnetic permeability of vacuum and the spontaneous magnetization%
, respectively. Furthermore, \(\BM\) is a unit-vector field that encodes the orientation of magnetization, such that \( \BM \in \CS^2\), where \( \CS^{d-1} = \{ \Bx \in \SR^d ~|~ \Bx \cdot \Bx = 1 \}\) is the $d$$-$$1$-dimensional unit sphere embedded in \(\SR^d\). The corresponding static Landau-Lifshitz equation will be derived as an Euler-Lagrange equation of the variational formulation to be discussed next.

\subsection{Variational formulation of the micromagnetic problem}
\subsubsection{Variational principle and Euler-Lagrange equations}
As independent field variables of the micromagnetic problem we consider the scalar magnetic potential~\(\Phi\) and the magnetization director~\(\BM\) as
\begin{equation}
	\Phi:
	\left\{
	\begin{array}{l}
		\OM \rightarrow \SR \\
		\Bx \mapsto \Phi(\Bx)
	\end{array}
	\right.
	\quad \mbox{and} \quad
	\BM:
	\left\{
	\begin{array}{l}
		\CB \rightarrow \CS^{d-1} \\
		\Bx \mapsto \BM(\Bx)
	\end{array}
	\right. .
\end{equation}
Based on the magnetic potential, we define the magnetic field \(\BH = -\grad \Phi\), such that the Maxwell equation \cref{eq:maxwellmagnet}$_2$ is identically satisfied.

In absence of externally applied fields and assuming isotropic behavior, the micromagnetic energy reads
\begin{align}
	\PI(\Phi, \BM) = \int_\CB \psi_\text{ex}(\grad \BM)\dV+ \int_\OM \psi_\text{stray}(\grad \Phi,\BM)  \dV,
	\label{eq:totalEnergyGeneric}
\end{align}
where \(\psi_\text{ex}\) is the exchange-energy density and \(\psi_\text{stray}\) is the stray-field (or demagnetization) energy density. They are given by \cite{miehe2012geometrically,difrattaVariationalPrinciplesMicromagnetics2019,keipVariationallyConsistentPhasefield2019a,reichel2022comparative,Wulfinghoff2022}
\begin{equation}
	\begin{aligned}
		\psi_\text{ex}(\grad \BM)         & =\frac{A}{2}\|\grad\BM\|^2 ,                                                   \\
		\psi_\text{stray}(\grad \Phi,\BM) & =  \mu_0 \widehat{M}_s \BM \cdot \grad \Phi -\frac{\mu_0}{2} \|\grad \Phi\|^2,
	\end{aligned} \label{eq:totalEnergyFirst}
\end{equation}
where \(A\) is the exchange-energy coefficient and 
\begin{align}
    \widehat{M}_s(\Bx) = \begin{cases}
        M_s, &\Bx \in \CB, \\
        0, &\text{otherwise}.
    \end{cases}\label{eq:magneticsaturation_def}
\end{align}

For a simpler notation, we introduce a dimensionless version of the total energy. In that consequence, we also change the definition of the coordinates $L\Bx\rightarrow \Bx$, where the length \(L =  \ell_{ex}/\sqrt{2}\) is defined through the magnetostatic exchange length \(\ell_{ex} = \sqrt{2A/(\mu_0 M_s^2)}\)~\cite{aboDefinitionMagneticExchange2013}. The dimensionless version of the total energy is accomplished by using the relations
\begin{equation}
	\phi(L\Bx) = \frac{\phi(\Bx)}{M_s L},
	\quad
	\BM(L\Bx)=\Bm(\Bx),
	\quad
	\PI(\Phi,\BM) = \widehat{\PI}(\phi,\Bm) \frac{L^2}{A},
	\quad
	\BH(L\Bx) = \frac{\Bh(\Bx)}{M_s},
	\quad
	\widehat{M}_s = \widehat{m}_s M_s ,
	\label{eq:nondim}
\end{equation}
In what follows, we stick to the notation '$\curl$', '$\divL$' and '$\grad$' for the differential operators, although from now on they are related to the dimensionless coordinates $\Bx$. The dimensionless total energy then reads
\begin{align}
	\widehat{\PI}(\phi,\Bm) = \int_{\widehat{\CB}} \frac{1}{2}\|\grad\Bm\|^2 + \grad \phi \cdot ( \widehat{m}_s \Bm ) \dv- \int_{\widehat{\OM}}   \frac{1}{2}\|\grad \phi\|^2 \dv
	\label{eq:totalEnergyDimless}
\end{align}
where \(\widehat{\OM}\) and \(\widehat{\CB}\) denote the appropriately rescaled domains. The corresponding variational principle is given as
\begin{align}
\boxed{
\{\phi^\ast, \Bm^\ast \} = \arg \left\{\sup_{\phi \in W^{k,p}(\widehat{\OM},\SR)} \inf_{\vphantom{\textnormal{\Large X}}\Bm \in W^{l,q}(\widehat{\OM},\CS^{d-1})}  \PI(\phi, \Bm) \right\}
}
\label{eq:totalEnergyVar}
\end{align}
where \(W^{k,p}\) denotes a Sobolev space. The corresponding Euler-Lagrange equations read
\begin{alignat}{6}
	\updelta  \phi: &  & \quad  \divL (-\grad \phi +  \Bm)  \,              &  & =       0 &  & \text{ in }    & \widehat{\CB}                     & \text{ in } \SR,  \label{eq:ELVecPotMag1}   \\
	\updelta \phi:  &  & \quad  \divL (- \grad \phi )  \,                   &  & =       0 &  & \text{ in } \, & \widehat{\OM} \setminus \widehat{\CB} & \text{ in } \SR,  \label{eq:ELVecPotMag1_1} \\
	\updelta \phi:  &  & \quad  (-\grad \phi+ \Bm) \cdot  \Bn^\CB\,         &  & =       0 &  & \text{ on } \, & \pt \widehat{\CB}                 & \text{ in } \SR,  \label{eq:ELVecPotMag2}   \\
	\updelta \phi:  &  & \quad (-\grad \phi) \cdot \Bn^\OM      \,          &  & =       0 &  & \text{ on } \, & \pt \widehat{\OM}                 & \text{ in } \SR,  \label{eq:ELVecPotMag2_1} \\
	\updelta \Bm:   &  & \quad  \Bm \times(\divL(\grad\Bm) + \grad \phi )\, &  & = \Bzero  &  & \text{ in } \, & \widehat{\CB}                     & \text{ in } \SR^3,\label{eq:ELVecPotMag3}   \\
	\updelta \Bm:   &  & \quad  \Bm \times (\grad\Bm \,\Bn)\,               &  & = \Bzero  &  & \text{ on } \, & \pt \widehat{\CB}                 & \text{ in } \SR^3,\label{eq:ELVecPotMag4}
\end{alignat}
where \(\Bn^\CB\) and \(\Bn^\OM\) denote the unit outward normal vectors on \(\pt \CB\) and \(\pt\OM\), respectively. The first two equations correspond to Gauss's law of magnetostatics for magnetic and non-magnetic matter, respectively, see \cref{eq:maxwellmagnet}. The third and fourth equation describe associated boundary conditions. The last two equations are related to the conformation of the magnetization, which needs to adhere the unit-sphere constraint $\| \Bm \| = 1$. The latter is usually accomplished by defining its variation as \(\updelta\Bm = \updelta\Bom \times \Bm\), where \(\updelta\Bom \) is a virtual axial vector of rotation, here given by the virtual spin of the magnetization. This then results in the usual static form of the Landau-Lifshitz equation \(\Bm \times \Bh_{\Re\Rf\Rf}= \Bzero\). Therefore, the variation of the magnetization $\updelta \Bm$ is constrained to lie in the tangent bundle \( T \CS^{d-1} \) of the unit sphere $\CS^{d-1}$ related to $\Bm$~\cite{miehe2012geometrically}. Consequently, the corresponding Euler-Lagrange equations need to be fulfilled only in the tangent bundle. Therefore, and for the sake of brevity, we omit the usual notation \(\Bm \times \ldots\) of this constraint and simply demand that the corresponding Euler-Lagrange equations
\begin{alignat}{5}
	\updelta\Bm: &  & ~ \divL(\grad\Bm) + \grad \phi     \, & = \Bzero &  & \cmc{ \text{ in }} & \widehat{\CB}\text{ in } T \CS^{2},  \label{eq:ELVecPotMag3_Mod}   \\
	\updelta\Bm: &  & ~ \quad   (\grad\Bm)  \, \Bn          & = \Bzero &  & \cmc{ \text{ on }} & \pt\widehat{\CB}\text{ in } T \CS^{2}, \label{eq:ELVecPotMag4_Mod}
\end{alignat}
have to be fulfilled in the tangent bundle of \(\CS^2\).

\cref{eq:ELVecPotMag2,eq:ELVecPotMag2_1} can be recast using \cref{eq:magneticsaturation_def} at the common surface of \(\widehat{\CB}\) and the free space \(\widehat{\OM}\setminus\widehat{\CB}\) as the jump condition
\begin{align}
\jumpl -\grad \phi + \widehat{m}_s\Bm \jumpr \cdot  \Bn=0
\ \text{ in } \
\SR,
\label{eq:jump}
\end{align}
where \(\Bn\) can be either \(\Bn^{\widehat{\CB}}\) or \(\Bn^{\widehat{\OM} \setminus \widehat{\CB}}\) since \(\Bn^{\widehat{\CB}}= -\Bn^{\widehat{\OM} \setminus \widehat{\CB}}\). Above, \(\jumpl(\cdot )\jumpr \) denotes the jump across the interface separating the body from the surrounding free space.

From a given magnetization, the scalar potential can be directly calculated. It is the solution of \cref{eq:ELVecPotMag1,eq:ELVecPotMag1_1,eq:ELVecPotMag2,eq:ELVecPotMag2_1}.
This solution for the scalar potential \(\phi\) is also denoted as the fundamental solution of Laplace's equation in \(\SR^3\), see~\cite{difrattaVariationalPrinciplesMicromagnetics2019}, since it boils down to \(\Delta \phi= \widehat{m}_s\divL \Bm\).\footnote{Strictly speaking \(\Delta \phi= \widehat{m}_s\divL \Bm\) is a Poisson equation since the right-hand side is non-zero.} It is given as
\begin{align}
	\phi(\Bx) = \frac{1}{4\pi} \left(-\int_{\widehat{\CB}'} \frac{\divL'  \Bm(\Bx')}{\|\Bx-\Bx'\|_2 } \dv' + \int_{\pt\widehat{\CB}'} \frac{ \Bm(\Bx') \cdot \Bn(\Bx')}{\|\Bx-\Bx'\|_2 } \da'\right),
	\label{eq:scalarPotSol}
\end{align}
where \(\divL'\) denotes the divergence w.r.t. \(\Bx' \).
For a source of \cref{eq:scalarPotSol}, see~\cite{difrattaVariationalPrinciplesMicromagnetics2019},\cite[Eq. 6.3.48]{aharoniIntroductionTheoryFerromagnetism1996} or~\cite[Eq. 3.46]{bertottiMaxwellEquationsMagnetic1998}.

The stray-field energy can then be calculated as
\begin{align}
	\widehat{\PI}_\text{stray}= \int_{\widehat{\CB}}\widehat{\psi}_\text{stray}( \phi,\Bm)  \dv= \frac{1}{2} \left(\int_{\widehat{\CB}} \rho_v \phi_v+2\rho_v \phi_a \dv+ \int_{\pt\widehat{\CB}} \sigma_s \phi_a \da\right) ,
	\label{eq:stray}
\end{align}
see~\cite[Eq. 5]{nonakaMagnetostaticEnergyMagnetic1985}. Here, \(\rho_v=-\divL \Bm\) and  \(\sigma_s= \Bm\cdot \Bn\).  These two quantities are denoted as \emph{volume charges} and \emph{surface charges}. Additionally, \(\phi_v\) denotes the volume contribution in \cref{eq:scalarPotSol} and  \(\phi_a\) denotes the surface contribution.

Thus, we have
\begin{equation}
	\begin{aligned}
		{\widehat{\PI}}_\text{stray}= \frac{1}{8\pi} \Biggl( & \int_{\widehat{\CB}} \int_{\widehat{\CB}'}  \frac{\divL \Bm(\Bx) \divL'  \Bm(\Bx')}{\|\Bx-\Bx'\|_2 } \dv'  \dv 
\\
		-                                                & 2\int_{\pt\widehat{\CB}}\int_{\widehat{\CB}'} \frac{\divL' \Bm(\Bx') \Bm(\Bx) \cdot \Bn(\Bx)}{\|\Bx-\Bx'\|_2 } \dv' \da+ \int_{\pt\widehat{\CB}}\int_{\pt\widehat{\CB}'} \frac{\Bm(\Bx) \cdot \Bn(\Bx) \Bm(\Bx') \cdot \Bn(\Bx')}{\|\Bx-\Bx'\|_2 } \da' \da \Biggr),
	\end{aligned} \label{eq:straysol}
\end{equation}
where the double integral reflects the highly non-local nature of the stray-field energy. Using the divergence theorem  in the form \(\int g \divL \Bf \dv = \int g \Bf \cdot \Bn \da - \int \Bf \cdot (\grad g) \dv \)
yields the following representation of the stray energy
\begin{equation}
	\begin{aligned}
		{\widehat{\PI}}_\text{stray} & =
		\frac{1}{8\pi} \Biggl(
		\int_{\widehat{\CB}} \int_{\widehat{\CB}'} \Bm(\Bx)\cdot (\Bx-\Bx') \frac{\divL'  \Bm(\Bx')}{\|\Bx-\Bx'\|_2^3 } \dv'   -  \int_{\pt\widehat{\CB}'} \Bm(\Bx)\cdot (\Bx-\Bx')\frac{ \Bm(\Bx') \cdot \Bn(\Bx')}{\|\Bx-\Bx'\|_2^3 } \da' \dv\Biggr),
	\end{aligned} \label{eq:straysolTest}
\end{equation}
which can be compared to \cref{eq:scalarPotSol} and then translated to the well-known energy expression
\begin{equation}
	\begin{aligned}
		{\widehat{\PI}}_\text{stray} & = \frac{1}{2}  \int_{\widehat{\CB}} \grad \phi(\Bm) \cdot \Bm \dv =  -\frac{1}{2}  \int_{\widehat{\CB}} \Bh(\Bm) \cdot \Bm \dv.
	\end{aligned} \label{eq:straysolTest2}
\end{equation}
The variation of the stray energy is
\begin{equation}
	\begin{aligned}
		\updelta{\widehat{\PI}}_\text{stray} & = -\frac{1}{2}  \int_{\widehat{\CB}} \updelta\Bh(\Bm) \cdot \Bm+\Bh(\Bm) \cdot \updelta\Bm  \dv
	\end{aligned} \label{eq:straysolVariation_1}
\end{equation}
and the first part of this variation reads
\begin{equation}
	\begin{aligned}
		\updelta\Bh(\Bm) & =  \frac{1}{4\pi} \Biggl(
		\int_{\widehat{\CB}'}  (\Bx-\Bx') \frac{\divL'  \updelta\Bm(\Bx')}{\|\Bx-\Bx'\|_2^3 } \dv'   -  \int_{\pt\widehat{\CB}'} (\Bx-\Bx')\frac{ \updelta\Bm(\Bx') \cdot \Bn(\Bx')}{\|\Bx-\Bx'\|_2^3 } \da'\Biggr).
	\end{aligned} \label{eq:straysolVariationPart_1}
\end{equation}
Exchanging the variables \(\Bx\) and \(\Bx'\) introduces a minus sign, so that we have \(\updelta\Bh(\Bm)=\Bh(\Bm) \cdot \updelta\Bm\). This is also a direct consequence of the \emph{reciprocity theorem}, see~\cite[Eq. 3-48]{brown1963micromagnetics}.
The variation simplifies to
\begin{equation}
	\begin{aligned}
		\updelta{\widehat{\PI}}_\text{stray} & = - \int_{\widehat{\CB}} \Bh(\Bm) \cdot \updelta\Bm  \dv.
	\end{aligned} \label{eq:straysolVariation_2}
\end{equation}
Thus, we have shown that the Euler-Lagrange equations are indeed the same as in \cref{eq:ELVecPotMag3}. We therefore have now
\begin{alignat}{5}
	\updelta\Bm: &  & \quad  \divL(\grad\Bm) + \grad \phi(\Bm)    \, & = \Bzero &  & \cmc{\text{ in }} & \widehat{\CB}\text{ in }    T \CS^{d-1}, \label{eq:ELVecPotMag_onlyM0} \\
	\updelta\Bm: &  & \quad  (\grad\Bm) \,\Bn                          & = \Bzero &  & \cmc{\text{ on }} & \pt\widehat{\CB}\text{ in } T \CS^{d-1}. \label{eq:ELVecPotMag_onlyM1}
\end{alignat}
In the following, these definitions are specified to the case of a nanodot.

\subsubsection{Application to nanodots}

The Euler-Lagrange equations derived above are valid for any shape of $\CB$ and can usually not be solved analytically. As will be shown next, it is however possible to derive accurate, semi-analytical solutions for the magnetization field of cylindrical specimens such as magnetic nanodots.

\paragraph{Curling mode.}
\label{sec:curling_mode}

For certain specimen shapes, the energetic minimum is represented by a single, rotationally symmetric vortex state. In the following, we present an ansatz that is tailor-made to reproduce this phenomenon.
It is encoded as
\begin{align}
	\Bm= \begin{bmatrix} -\sin \theta \sin f(\rho) \\ \cos \theta \sin f(\rho) \\ \cos f(\rho)  \end{bmatrix} .
	\label{eq:nanodotAnsatz}
\end{align}
Here, \(\rho \in [0,\widehat{R}] \)  is the radial coordinate and \(\theta \in [0, 2 \pi]\) is the angular coordinate. The dimensionless radius of the nanodot specimen is denoted by \(\widehat{R}\). We will denote the dimensionless thickness of the cylinder as \(\widehat{H}\) and later also use \(\delta=\widehat{H}/2\). The vector components in \cref{eq:nanodotAnsatz} refer to the usual Cartesian coordinates \(x,y\) and \(z\). Furthermore, \(f(\rho)\) is an initially unknown function for the rotationally symmetric ansatz that will be specified in \cref{sec:annihil}. The assumed slenderness of the nanodot also justifies the fact that the magnetization given in~\cref{eq:nanodotAnsatz} is constant across the thickness.

Introducing~\cref{eq:nanodotAnsatz} into the energy~\cref{eq:totalEnergyFirst}, the exchange energy is obtained as
\begin{align}
	\widehat{\PI}_\text{ex}^{\text{CM}} (\grad \Bm) = \int_{\widehat{\OM}} \frac{1}{2}\|\grad\Bm\|^2\dV = 2\pi \widehat{H}\int_0^{\widehat{R}} \frac{1}{2\rho}\left[{\left(\fracpt{f(\rho)}{\rho} \right)}^2\rho^2 + \sin^2 f(\rho)  \right]\,\Rd \rho ,
	\label{eq:totalEnergyExchange}
\end{align}
where the superscript 'CM' stands for 'curling mode'. The latter representation of \(\widehat{\PI}_\text{ex}\) can also be found in, e.g., \cite{brown1963micromagnetics,usovMagnetizationCurlingFine1993}.
Since \(\widehat{\PI}_\text{ex}\) is independent of the coordinates \(z\) and \(\theta\), preintegration in these directions was carried out, reducing the integral to only one dimension.

We now use \cref{eq:straysol} to calculate the stray-field energy.
Due to the specific form of the magnetization~\cref{eq:nanodotAnsatz}, we have \(\divL \Bm=0\), and are left with the double integral over the surfaces. Since the direction of magnetization according to~\cref{eq:nanodotAnsatz} is tangential to the lateral surface, we also have \(\Bm \cdot \Bn=0\).
Thus, only the contributions on the top surface \(\CB'^T\) and the bottom surface \(\CB'^B\) at \(z=\pm\delta=\pm\widehat{H}/2\) are nonzero.
At these surfaces we have \(\Bm \cdot \Bn=\pm m_z= \pm \cos f(\rho)\) due to \(\Bn={[0,0,\pm 1]}^T\). From this it follows
\begin{equation}
\begin{aligned}
\widehat{\PI}_\text{stray}^{\text{CM}}(\Bm)
& =
\frac{1}{8\pi} \int_{\pt\widehat{\CB}}\int_{\pt\widehat{\CB}'} \frac{\Bm(\Bx) \cdot \Bn(\Bx) \Bm(\Bx') \cdot \Bn(\Bx')}{\|\Bx-\Bx'\|_2 } \da' \da
\\
& =
\frac{1}{8\pi} \left(\int_{\pt\widehat{\CB}^T}\int_{\pt\widehat{\CB}'^T} \frac{\Bm_z(\rho,\theta,\delta)  \Bm_z(\rho',\theta',\delta) }{\|\Bx-\Bx'\|_2 } \da' \da \right.                  \\
& +
\left.\int_{\pt\widehat{\CB}^B}\int_{\pt\widehat{\CB}'^B} \frac{(-\Bm_z(\rho,\theta,-\delta))  (-\Bm_z(\rho',\theta',-\delta)) }{\|\Bx-\Bx'\|_2 } \da' \da \right.
\\
& +
2\left.\int_{\pt\widehat{\CB}^B}\int_{\pt\widehat{\CB}'^T} \frac{\Bm_z(\rho,\theta,\delta)  (-\Bm_z(\rho',\theta',-\delta)) }{\|\Bx-\Bx'\|_2 } \da' \da  \right)
\\
& =
\frac{1}{4\pi} \left(\int_0^{2\pi} \int_0^{\widehat{R}}\int_0^{2\pi} \int_0^{\widehat{R}}  \frac{\cos f(\rho)  \cos f(\rho') }{d } \rho' \,\Rd \rho' \,\Rd \theta'  \rho\, \Rd \rho \,\Rd \theta \right.
\\
& +
\left.\int_0^{2\pi} \int_0^{\widehat{R}}\int_0^{2\pi} \int_0^{\widehat{R}} \frac{-\cos f(\rho)  \cos f(\rho') }{\sqrt{d^2+4\delta^2} }  \rho' \,\Rd \rho' \,\Rd \theta'  \rho \,\Rd \rho \,\Rd \theta \right),
\end{aligned} \label{eq:totalEnergyStrayNanoDot}
\end{equation}
with \(d(\rho,\rho',\theta,\theta')^2=-2\rho\rho'\cos(\theta-\theta')+{(\rho')}^2+\rho^2\).
This is derived by inserting the definitions \(\Bx = {[\rho \sin \theta, \rho\cos \theta, \pm\delta]}^T\) and  \(\Bx' = {[\rho'\sin \theta', \rho'\cos \theta', \pm\delta]}^T\). Integration is accomplished with the usual Jacobian of polar coordinates, i.e.  \(\int \int (\cdot) \dx\dy = \int \int (\cdot) \rho \Rd \theta\Rd\rho \). 
 Since the denominator can be made independent of \(\theta\)\footnote{By realizing $\int_0^{2\pi}\int_0^{2\pi} \frac{1}{\sqrt{-2\rho\rho'\cos(\theta-\theta')+{(\rho')}^2+\rho^2}} \,\Rd \theta' \,\Rd \theta = \int_0^{2\pi}  \,\Rd \theta' 4 \int_0^{\pi/2} \frac{1}{\sqrt{-4\rho\rho'\sin^2(\theta)+({\rho'}+\rho)^2}} \,\Rd \theta $. }, similar to~\cite[Eq. 19]{aharoniMagnetostaticEnergyFerromagnetic1983}, one integration can be carried out directly and we get
\begin{equation}
	\begin{aligned}
		\widehat{\PI}_\text{stray}^{\text{CM}} (f) & = 2 \left(\int_0^{\pi/2} \int_0^{\widehat{R}} \int_0^{\widehat{R}}  \frac{\cos f(\rho)  \cos f(\rho') }{d } \rho' \,\Rd \rho'  \rho \,\Rd \rho \,\Rd \theta - \frac{\cos f(\rho) \cos f(\rho') }{\sqrt{d^2+4\delta^2} }  \rho' \,\Rd \rho' \rho \,\Rd \rho \,\Rd \theta \right),
	\end{aligned} \label{eq:totalEnergyStrayNanoDot2}
\end{equation}
with \(d(\rho,\rho',\theta)^2=-4 \rho\rho'\sin^2(\theta)+{(\rho'+\rho)}^2\).

Now, we can also integrate w.r.t.\ \(\theta\). Both integrals in \cref{eq:totalEnergyStrayNanoDot2} have the following form that can be integrated in terms of elliptic integrals,
\begin{align}
	\int_0^{\pi/2} \frac{\dx}{\sqrt{a-b\sin^2(x)}}  = \frac{1}{\sqrt{a}} \ellipticK\left(\frac{b}{a}\right),
	\label{eq:ellipticK}
\end{align}
where \(\ellipticK(m)\) is the \emph{complete elliptic integral of the first kind} with the parameter \(m=k^2\) and \(k\) is the \emph{elliptic modulus}. Thus, we can simplify \cref{eq:totalEnergyStrayNanoDot2} as
\begin{equation}
	\begin{aligned}
		\widehat{\PI}_\text{stray}^{\text{CM}} (f) & = 2  \int_0^{\widehat{R}} \cos f(\rho) \rho \int_0^{\widehat{R}} \cos f(\rho')\rho' \left(\frac{  1 }{\widehat{a} }\ellipticK\left(\frac{b}{\widehat{a}^2}\right) -  \frac{1 }{\sqrt{\widehat{a}^2+4\delta^2} }  \ellipticK\left(\frac{b}{\widehat{a}^2+4\delta^2}\right) \right) \Rd \rho'  \Rd \rho,
	\end{aligned} \label{eq:totalEnergyStrayNanoDot3}
\end{equation}
where we used \(b= 4\rho\rho' \) and \(\widehat{a}= \rho'+\rho\).
Now, we further introduce the shortcut \(\CK\) for the symmetric kernel
\begin{align}
	\CK(\rho,\rho') = \frac{  \rho\rho'  }{\widehat{a} }\ellipticK\left(\frac{b}{\widehat{a}^2}\right) -  \frac{\rho\rho'}{\sqrt{\widehat{a}^2+4\delta^2} }  \ellipticK\left(\frac{b}{\widehat{a}^2+4\delta^2}\right).
	\label{eq:kernel}
\end{align}
Finally, since \(\widehat{\PI}_\text{stray}^{\text{CM}} (f)\) is symmetric w.r.t.\ \(\rho\) and \(\rho'\), we transform the integral from the quadrilateral domain \([0,\widehat{R}]\times [0,\widehat{R}]\) to an integral over the triangular domain \(\rho \in [0,\widehat{R}], \rho' \in [0,\rho]\) by multiplying with 2 and changing the interval of integration accordingly. This yields
\begin{equation}
	\begin{aligned}
		\widehat{\PI}_\text{stray}^{\text{CM}} (f) & = 4  \int_0^{\widehat{R}} \cos f(\rho)  \int_0^{\rho} \cos f(\rho')\CK(\rho,\rho')  \Rd \rho'  \Rd \rho.
	\end{aligned} \label{eq:totalEnergyStrayNanoDot4}
\end{equation}
The total energy, only in terms of \(f\), is
\begin{align}
	\widehat{\PI}^{\text{CM}}(f) = 4  \int_0^{\widehat{R}} \cos f(\rho)  \int_0^{\rho} \cos f(\rho')\CK(\rho,\rho') \, \Rd \rho' + 2\pi \widehat{H}\frac{1}{2\rho}\left[{\left(\fracpt{f(\rho)}{\rho} \right)}^2\rho^2 + \sin^2 f(\rho)  \right]\,\Rd \rho.
	\label{eq:energy_f_nanodot}
\end{align}
This gives rise to the minimization principle
\begin{equation}
\begin{aligned}
\boxed{
f^\ast  = \arg \left\{ \inf_{f \in W^{k,p}(\CP,\SR)} \widehat{\PI}^{\text{CM}} (f) \right\}    
}
\end{aligned}
\label{eq:totalEnergyVarPrinciple}
\end{equation}
with $\widehat{\PI}^{\text{CM}} (f) = \widehat{\PI}^{\text{CM}}_\text{ex}(f) + \widehat{\PI}^{\text{CM}}_\text{stray}(f)$ and $\CP = [0,\widehat{R}]$.

The first variation of the individual energy contributions reads
\begin{equation}
	\begin{aligned}
		\de\widehat{\PI}_\text{stray}^{\text{CM}} (f) & = -8  \int_0^{\widehat{R}} \sin f(\rho)  \int_0^{\rho} \cos f(\rho') \CK(\rho,\rho')  \Rd \rho'  \de f(\rho)\Rd \rho
	\end{aligned} \label{eq:ELCurlingMode_Stray}
\end{equation}
and
\begin{equation}
	\begin{aligned}
		\updelta\widehat{\PI}_\text{ex}^{\text{CM}} (f) & =  2\pi \widehat{H} \int_0^{\widehat{R}}\left(\frac{\sin f(\rho)\cos f(\rho)}{\rho}-\fracpt{f(\rho)}{\rho}  -\fracpt{^2f(\rho)}{\rho^2} \rho \right) \de f(\rho)\Rd \rho.
	\end{aligned} \label{eq:ELCurlingMode_EX}
\end{equation}
The corresponding Euler-Lagrange equation is a non-autonomous, nonlinear second order \emph{integro-differential} equation that reads
\begin{equation}
	\begin{aligned}
		\updelta & f:   \sin f(\rho) \left( -4\int_0^{\widehat{R}} \cos f(\rho') \CK(\rho,\rho') \, \Rd \rho'+2\pi \widehat{H}\frac{\cos f(\rho)}{\rho}\right) - 2\pi \widehat{H}\left(\fracpt{f(\rho)}{\rho}  +\fracpt{^2f(\rho)}{\rho^2}\rho\right)=0 \quad &  & \forall \rho \in \CP,       \\
		\updelta & f:   \fracpt{f(\rho)}{\rho}=0 \quad                                                                                                                                                                                 &  & \text{at }  \rho = \widehat{R}. \\
	\end{aligned} \label{eq:ELCurlingMode}
\end{equation}
Here, the interval of the inner integral is again \(\widehat{R}\) instead of \(\rho\), since \cref{eq:ELCurlingMode} is not symmetric in \(\rho\) and \(\rho'\) anymore.
This means that the variation of the demagnetization contains the factor \(4\) instead of \(8\). Parts of this ordinary differential equation can also be found in the literature, specifically in~\cite[p. 97, eq. 6--19]{brown1963micromagnetics}.
In the latter reference, the stray energy is missing, but anisotropic effects and an external applied field are taken into account. We are not aware of a closed-form solution of the derived~\cref{eq:ELCurlingMode}.

\paragraph{In-plane single-domain state.}
\label{sec:in-plane_section} 
The in-plane single-domain state of the nanodot is encoded in a homogeneous magnetization throughout the nanodot.
Note that this is not a realistic assumption for cylindrical nanodots, since a perfectly homogeneous magnetization can only be observed in ellipsoidal bodies~\cite{PhysRev.67.351}. Nevertheless, we assume that for a thin specimen it is a reasonable approximation, see \cite{kobayashiSurfaceMagnetic1992,wysinMagneticExcitations2015} and \cite[Ch. 6.1.2]{aharoniIntroductionTheoryFerromagnetism1996}
for reference. Due to isotropy, we could assume any orientation of magnetization in the \(x\)-\(y\)-plane. In what follows, we choose \(\Bm= \Be_x\). Due to homogeneity, we further have \(\grad \Bm =\Bzero \), so that the exchange energy vanishes identically and only the stray energy \cref{eq:straysol} has to be evaluated. Since \(\Bm\) is tangential to the top and bottom surface, we only have to evaluate \cref{eq:straysol} at the lateral surface of the cylinder.
Again, the volume terms vanish due to \( \divL \Bm=0\).
Hence, with \(\Bm \cdot \Bn = \cos \theta\) the stray energy at the lateral surface is
\begin{equation}
	\begin{aligned}
		\widehat{\PI}_\text{stray}^{\text{IP}} & = \frac{1}{8\pi} \int_{\pt\widehat{\CB}}\int_{\pt\widehat{\CB}'} \frac{\Bm(\Bx) \cdot \Bn(\Bx) \Bm(\Bx') \cdot \Bn(\Bx')}{\|\Bx-\Bx'\|_2 }  \da' \da                                                                                                          \\
		                                   & =  \frac{1}{8\pi} \int_{-\delta}^\delta   \int_{0}^{2\pi} \int_{-\delta}^\delta \int_{0}^{2\pi} \frac{\cos(\theta) \cos(\theta')}{\sqrt{{(z-z')}^2-2 \widehat{R}^2 \cos(\theta-\theta')+2 \widehat{R}^2 }} \widehat{R}^2\,\Rd z' \,\Rd\theta'\,\Rd\theta \,\Rd z,
	\end{aligned} \label{eq:totalEnergyStrayNanoDot_INPLANE}
\end{equation}
where 'IP' stands for 'in plane'. Furthermore, \(\Bx = [\widehat{R}\sin \theta, \widehat{R}\cos \theta, z]^T\), \(\Bx' = [\widehat{R}\sin \theta', \widehat{R}\cos \theta', z']^T\), and the infinitesimal area element is \(\|\frac{\partial \Bn}{\partial\theta }\times \frac{\partial \Bn}{\partial z}\|_2 \,\Rd\theta \,\Rd z =\widehat{R} \,\Rd\theta \,\Rd z\). Additionally, for the transformation of the denominator, we used \(\sin(\theta')\sin(\theta) +\cos(\theta')\cos(\theta)=\cos(\theta'-\theta)\).

The integrals w.r.t.\ the thickness coordinates \(z\) and \(z'\) can be evaluated analytically. Thus, we get
\begin{equation}
	\begin{aligned}
		\widehat{\PI}_\text{stray}^{\text{IP}} & =  -\frac{\widehat{R}^2}{4\pi}    \int_{0}^{2\pi} \int_{0}^{2\pi}\cos\theta \cos \theta' (\delta\ln(-2\delta + q) - \delta\ln(2\delta + q) + q - \sqrt{-4\delta^2 + q^2}) \,\Rd\theta'\,\Rd\theta ,
	\end{aligned} \label{eq:totalEnergyStrayNanoDot_INPLANE2}
\end{equation}
with \(q^2= 4\delta^2 + 2\widehat{R}^2(1-\cos(\theta - \theta')) \). The single integration w.r.t.\ \(z'\) was also shown in~\cite{caciagliExactExpressionMagnetic2018}. However, in contrast to our case, the authors of~\cite{caciagliExactExpressionMagnetic2018} derived \(\phi(\Bx)\) by integrating the equivalent of \cref{eq:scalarPotSol} w.r.t.\ \(z'\) and \(\theta' \).

Since we are interested in the energy, two further integrations remain.

We start with the integration w.r.t. \(\theta'\).
First, we perform a substitution of variables by using \(x=\pi-(\theta'-\theta)\) and get

\begin{equation}
\begin{aligned}
\widehat{\PI}_\text{stray}^{\text{IP}} & = -\frac{\widehat{R}^2}{4\pi}\int_{0}^{2\pi}\cos\theta\int_{-\pi+\theta}^{\pi+\theta}\cos(\theta-x)\Bigg(\underbrace{\ln(2\delta+\sqrt{2\widehat{R}^2\cos x+2\widehat{R}^2+4\delta^2})\delta}_{(1)}
\\
&
\underbrace{-\ln(-2\delta+\sqrt{2\widehat{R}^2\cos x+2\widehat{R}^2+4\delta^2})\delta }_{(2)}
\underbrace{-\sqrt{2\widehat{R}^2\cos(x)+2\widehat{R}^2+4\delta^2}}_{(3)}\underbrace{+\sqrt{2}\widehat{R}\sqrt{\cos x+1}}_{(4)}\Bigg) \dx\,\Rd\theta,
\end{aligned} \label{eq:totalEnergyStrayNanoDot_INPLANE_thirdIntegration}
\end{equation}
which yields a better separation of variables between \(x\) and \(\theta\). Next, we evaluate this integral part by part, starting with the fourth underbraced term.
Here, direct integration succeeds with the computer algebra system Maple~\cite{maple},
\begin{equation}
	\begin{aligned}
		-\frac{\widehat{R}^2}{4\pi}\int_{0}^{2\pi}\cos\theta\int_{-\pi+\theta}^{\pi+\theta}\cos(\theta-x)\sqrt{2}\widehat{R}\sqrt{\cos x+1} \dx\,\Rd\theta = -\frac{2}{3\pi} \widehat{R}^3\int_{0}^{2\pi} \cos^2 \theta \,\Rd\theta = -\frac{2}{3}\widehat{R}^3.
	\end{aligned} \label{eq:totalEnergyStrayNanoDot_INPLANE_thirdIntegration_part4}
\end{equation}
The third term can also be calculated with Maple as
\begin{equation}
	\begin{aligned}
		\frac{\widehat{R}^2}{4\pi}\int_{0}^{2\pi}\cos\theta\int_{-\pi+\theta}^{\pi+\theta}\cos(\theta-x)\sqrt{2\widehat{R}^2\cos(x)+2\widehat{R}^2+4\delta^2} \dx\,\Rd\theta \\
		= \frac{2}{3\pi}\left((R^2+2\delta^2)\ellipticE(-\beta^2)-2\ellipticK(-\beta^2)(R^2+\delta^2)\right)\cos^2\theta\delta,
	\end{aligned} \label{eq:totalEnergyStrayNanoDot_INPLANE_thirdIntegration_part3}
\end{equation}
where we introduced the factor of slenderness \(\beta = \widehat{R}/\delta\). Additionally, we used
\begin{align}
	\ellipticE(m) = \int_0^{\pi/2} \sqrt{1-m\sin^2(x)}\dx ,
	\label{eq:ellipticE}
\end{align}
where \(\ellipticE\) is called \emph{complete elliptic integral of the second kind}. For the evaluation of the second term in \cref{eq:totalEnergyStrayNanoDot_INPLANE_thirdIntegration}, we need to apply integration by parts, where we use \(\cos(\theta-x)\) as one factor, which can be integrated in a straight-forward manner. This yields
\begin{equation}
	\begin{aligned}
		\frac{\widehat{R}^2}{4\pi}\int_{0}^{2\pi}\cos\theta\int_{-\pi+\theta}^{\pi+\theta}\cos(\theta-x)\ln(-2\delta+d(x))\delta  \dx\,\Rd\theta =   -\int_{0}^{2\pi}\int_{-\pi+\theta}^{\pi+\theta}\frac{\cos(\theta)\widehat{R}^4\sin(\theta-x)\sin x \delta}{4\pi d(x)\left(-2\delta+d(x)\right)} \dx\,\Rd\theta,
	\end{aligned} \label{eq:totalEnergyStrayNanoDot_INPLANE_thirdIntegration_part2}
\end{equation}
with $d(x)=\sqrt{2\widehat{R}^2\cos x+2\widehat{R}^2+4\delta^2}$.
Here, the boundary terms vanish, since \(\sin(-(-\pi + \theta)+\theta)=0\) and \(\sin(-(\pi + \theta)+\theta)=0\).

Now, by expanding the numerator with the angle difference identity \(\sin(\theta - x) = \sin \theta \cos x - \cos \theta \sin x \), we arrive at two terms that have separated variables
\begin{equation}
	\begin{aligned}
		-\int_{0}^{2\pi}\int_{-\pi+\theta}^{\pi+\theta}\frac{\cos\theta\widehat{R}^4\sin(\theta-x)\sin x \delta}{4\pi d(x)(-2\delta+d(x))} \dx\,\Rd\theta  =-\frac{\widehat{R}^4\delta}{4\pi}\int_{0}^{2\pi}\int_{-\pi+\theta}^{\pi+\theta}\frac{\cos\theta\sin \theta \cos x\sin x - \cos^2\theta \sin^2 x }{d(x)\left(-2\delta+d(x)\right)} \dx\,\Rd\theta.
	\end{aligned} \label{eq:totalEnergyStrayNanoDot_INPLANE_thirdIntegration_part2_1}
\end{equation}
The first term in the numerator vanishes, since the integral of this part is only a function \(\cos(x)\) which yields the same value for \(x= -\pi+\theta \) and \(x=\pi+\theta\). The second part however is non-zero and yields, again evaluated with Maple,
\begin{equation}
	\begin{aligned}
		 & \frac{\widehat{R}^4\delta}{4\pi}\int_{0}^{2\pi}\int_{-\pi+\theta}^{\pi+\theta}\frac{\cos^2 \theta \sin^2 x }{d(x)\left(-2\delta+d(x)\right)} \dx\,\Rd\theta  =\frac{\delta}{4\pi} (4 (\widehat{R}^2+\delta^2)\ellipticK(-\beta^2)-4\ellipticE(-\beta^2)\delta^2+\pi \widehat{R}^2) \int_{0}^{2\pi} \cos^2\theta \,\Rd\theta \\
		 & =\delta \left( (\widehat{R}^2+\delta^2)\ellipticK(-\beta^2)-\ellipticE(-\beta^2)\delta^2+\frac{\pi \widehat{R}^2}{4}\right) .
	\end{aligned} \label{eq:totalEnergyStrayNanoDot_INPLANE_thirdIntegration_part2_2}
\end{equation}
The first term can be evaluated similarly to the second one, i.e., via integration by parts and by realizing the vanishing boundary terms. Then, expanding the integral yields two terms. After integration, one of these terms is a function of \(\cos x\), so that it vanishes in consideration of the integration limits. The remaining term is again evaluated with the help of computer algebra~\cite{maple}, which yields
\begin{equation}
	\begin{aligned}
		 & -\frac{\widehat{R}^2}{4\pi}\int_{0}^{2\pi}\cos\theta\int_{-\pi+\theta}^{\pi+\theta}\cos(\theta-x)\ln(2\delta+\sqrt{2\widehat{R}^2\cos x+2\widehat{R}^2+4\delta^2})\delta \dx\,\Rd\theta \\
		 & =\frac{\delta}{4\pi} (4 (\widehat{R}^2+\delta^2)\ellipticK(-\beta^2)-4\ellipticE(-\beta^2)\delta^2-\pi \widehat{R}^2)\int_{0}^{2\pi} \cos^2\theta \,\Rd\theta                       \\
		 & =\delta\left((\widehat{R}^2+\delta^2)\ellipticK(-\beta^2) -\ellipticE(-\beta^2) \delta^2-\frac{\pi \widehat{R}^2 }{4} \right).
	\end{aligned} \label{eq:totalEnergyStrayNanoDot_INPLANE_thirdIntegration_part1}
\end{equation}
In total, evaluation of the stray energy, \cref{eq:totalEnergyStrayNanoDot_INPLANE}, provides
\begin{equation}
	\begin{aligned}
		\widehat{\PI}_\text{stray}^{\text{IP}} & =  -\frac{2}{3}( (\beta^3-\beta)\ellipticE(-\beta^2) - (\beta^3 + \beta)\ellipticK(-\beta^2) + 1 )R^3,
	\end{aligned} \label{eq:totalEnergyStrayNanoDot_final}
\end{equation}
which, to our best knowledge, cannot be found in this particular form in the literature. We would however like to mention the demagnetization energy derived in \cite{tandonComputationDemagnetizationTensor2004}, which has a different representation, but which gives numerically equivalent results. In contrast to \cref{{eq:totalEnergyStrayNanoDot_final}}, the representation in \cite{tandonComputationDemagnetizationTensor2004} involves hypergeometric functions but can be used for any homogeneous orientation of magnetization. Furthermore, there are certain similarities to the demagnetization factors calculated in \cite{Joseph1966} and to the scalar potential calculated in \cite{caciagliExactExpressionMagnetic2018}.

\paragraph{Out-of-plane single-domain state.}
\label{sec:out-of-plane_section}

The energy of the out-of-plane single-domain state can be directly derived from the curling mode~\cref{eq:nanodotAnsatz} by setting \(f(\rho)=0\). Then, we end up with the energy
\begin{align}
	\widehat{\PI}^{\text{OOP}}(f) = 4  \int_0^{\widehat{R}} \int_0^{\rho} \CK(\rho,\rho') \, \Rd \rho' \,\Rd \rho ,
	\label{eq:energy_f_nanodot_out-of-plane}
\end{align}
where 'OOP' stands for 'out of plane'. The first part of the kernel,~\cref{eq:kernel}, can be evaluated analytically,
\begin{align}
	\widehat{\PI}^{\text{OOP}}_\text{stray} = \frac{4 R^3}{3} - \int_0^{\widehat{R}} \int_0^{\rho} \frac{4\rho\rho'}{\sqrt{\widehat{a}^2+4\delta^2} }  \ellipticK\left(\frac{b}{\widehat{a}^2+4\delta^2}\right) \, \Rd \rho' \,\Rd \rho,
	\label{eq:energy_f_nanodot_out-of-plane2}
\end{align}
and the second part is left for numerical integration. In \cite{tandonComputationDemagnetizationTensor2004}, this term is derived analytically using hypergeometric functions,
\begin{align}
	\widehat{\PI}^{\text{OOP}}_\text{stray} = \frac{R^3}{6}\left[-4+3\pi\sqrt{\tau^2+1}\hypergeom{2}{1}\left(-\frac{1}{2}, \frac{3}{2};2;\frac{1}{\tau^2+1}\right)\right],
	\label{eq:energy_f_nanodot_out-of-plane3}
\end{align}
where \(\hypergeom{2}{1} \) is the hypergeometric function and $\tau=1/\beta$.

\section{Results and discussion}
\subsection{Profiles of the out-of-plane magnetization}
\label{sec:annihil}

\begin{figure}[t]
	\centering
	\includegraphics[page=1]{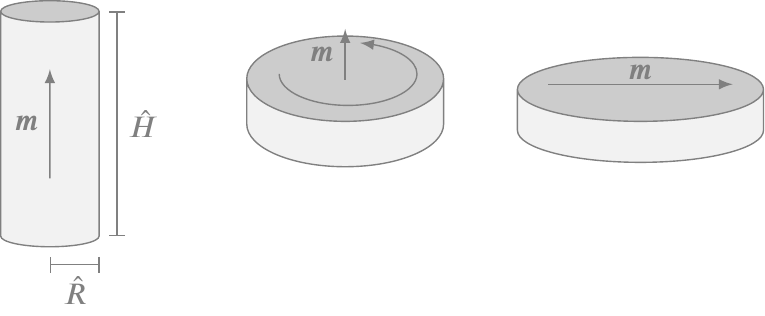}
	\caption{The three minimizing configurations depending on the thickness-to-radius ratio. On the left, the magnetization is homogeneously distributed throughout the specimen and points in the direction of the cylinder's axis. In the middle, the curling mode is shown with the magnetization pointing out of plane in the center of the vortex as discussed in~\cref{sec:curling_mode}. On the right, the homogeneous, in-plane magnetization as discussed in~\cref{sec:in-plane_section} is shown.}
	\label{fig:configurations}
\end{figure}

We now use \cref{eq:totalEnergyStrayNanoDot_final} and \cref{eq:energy_f_nanodot} to determine at what point the curling mode becomes unstable. For this, we have to take three different modes into account. These three modes are shown in \cref{fig:configurations}.
The left and the center one are entirely captured by the curling mode~\cref{eq:nanodotAnsatz}. For the right one, with in-plane constant magnetization, the magnetostatic energy was derived in \cref{sec:in-plane_section}. The (sudden) switch between these three states is of particular interest and arises as an instability phenomenon.

To study the minimizers of the curling mode, we use Ritz's method. This distinguishes our approach from classical models in the literature, from which many are based on the approach suggested by~\cite{usovMagnetizationCurlingFine1993} or slight modifications thereof. For completeness, we write down the ansatz for the out-of-plane magnetization given in~\cite{usovMagnetizationCurlingFine1993}
\begin{align}
	m_z(\rho)= \begin{cases} \sqrt{1-\frac{2a\rho}{a^2+\rho^2}}, & 0\leq \rho \leq a,       \\
              0,                                  & a\leq \rho \leq \widehat{R},
	           \end{cases}
	\label{eq:usov-ansatz}
\end{align}
where the size of the domain with non-zero out-of-plane magnetization \(a\) is derived by minimizing the energy w.r.t.\ \(a\). We note that the above function for the magnetization is non-differentiable at \(\rho=a\). In other words, it has a non-physical kink precisely at this position.

Since we want to arrive at the exact minimizers of \cref{eq:energy_f_nanodot}, at least asymptotically, we use a Fourier series expansion given by
\begin{align}
	\widehat{f}(\rho)= \sum_{i=1}^{n} a_i \sin\left(\rho \frac{(2i-1)\pi}{\widehat{R}/2}\right)
	\label{eq:ritz_fourier}
\end{align}
and assume that \(\widehat{f}(\rho) \in V \subset W^{k,p}(\CP,\SR)\). The series expansion includes only terms that satisfy the boundary conditions \(f(0)=0 \) and \(\fracpt{f(\rho)}{\rho}|_{\rho= \widehat{R}}=0\) for any \(a_i\).The first condition follows from circular symmetry, requiring the magnetization to point in \(z\)-direction at \(\rho=0\). The second condition follows from the Euler-Lagrange equation at the boundary, see \cref{eq:ELVecPotMag4}.

For the numerical solution of the nonlinear problem of energy minimization, the trust region method~\cite{conn2000trust} is used. The gradient and the Hessian of the energy w.r.t.\ \(a_i\) are calculated using automatic differentiation in our \Cpp-implementation using~\cite{Autodiff}.

\begin{figure}[t]
	\centering
	\includegraphics[page=1,width=0.6\textwidth]{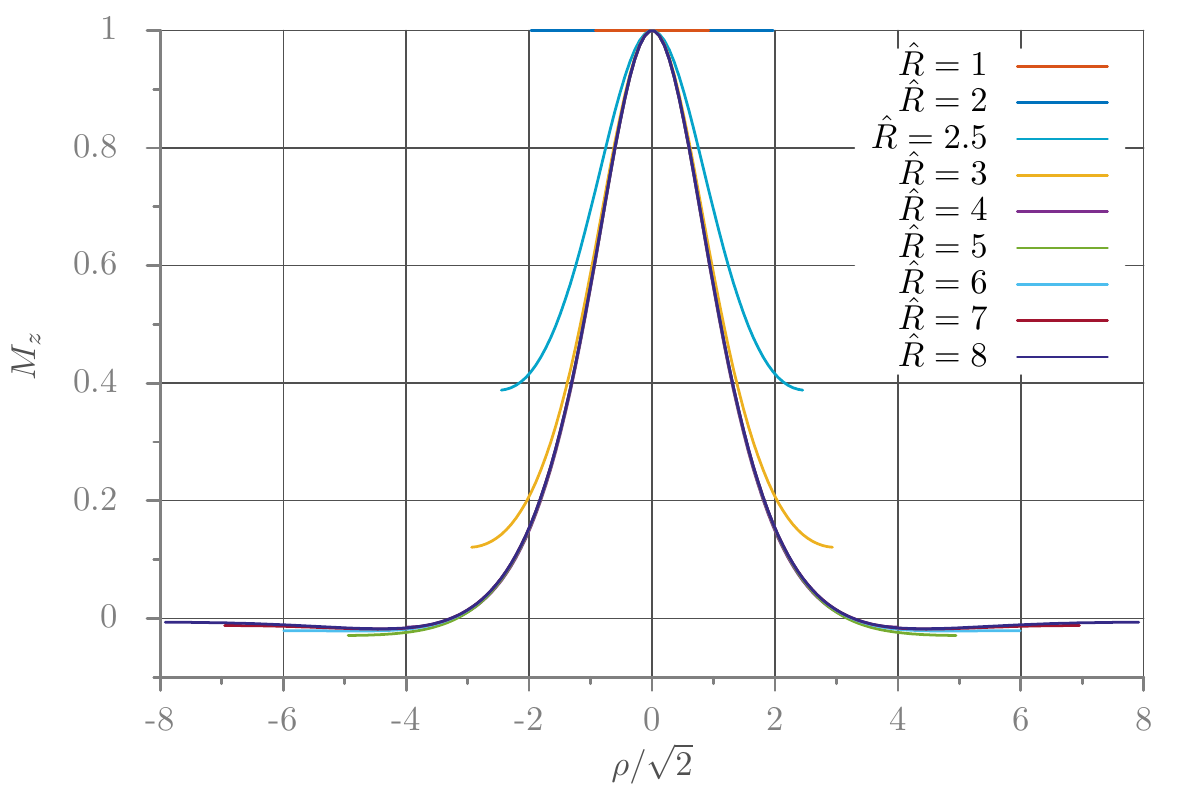}
	\caption{Profiles of the \(z\)-component of the magnetization \(\Bm\) for different nanodot radii and a fixed height \(\widehat{H}= 3\sqrt{2} \) at the minimizing energy configuration. \(\widehat{H}= 3\sqrt{2} \) corresponds to three times the exchange length \(\ell_\text{ex}\). The plots were obtained using \cref{eq:nanodotAnsatz}. The horizontal axis is scaled such that the unit is in terms of the exchange length \(\ell_\text{ex}\).}
	\label{fig:MZ_Fourier}
\end{figure}

The obtained profiles of the out-of-plane magnetization \(m_z = \cos f(\rho)\) are shown in \cref{fig:MZ_Fourier} for several nanodot radii and a fixed thickness \(\widehat{H}= 3\sqrt{2}\). For all computations, we increase the number of coefficients \(a_i\) until the change in the minimizing energy is below \(1e-6\). For reference, $\widehat{\PI}^{\text{CM}}$ varies from $33.54$ to $153.91$ for geometries $\{\widehat{R}; \widehat{H}\}$ given by $\{3.28; 5.95\}$ and $\{14.14; 14.14\}$, respectively, where the smaller energy is associated with the separated red dot in the phase diagram shown in~\cref{fig:Minimizer_of_modes_withData} (please refer to the zoomed region for a detailed view). For the raw data, see \cite{darus-3103_2023}.   The energy decrease of \(1e-6\) is usually captured with $i$ between 3 and 25. The results are in qualitative agreement with the results obtained from micromagnetic simulations documented in~\cite[Fig.~2]{guslienkoVortexStateStability2004} and~\cite[Fig. 3]{raabeMagnetizationPatternFerromagnetic2000}. In particular, the overshooting to the negative regime \(M_z<0\) in the periphery of the vortex is captured. Furthermore, the decay \(M_z(\widehat{R})\rightarrow 0\) for \(\widehat{R}\rightarrow \infty\) is foreseeable.

\subsection{Phase diagram}

\begin{figure}[t]
	\centering
	\includegraphics[page=1,width=0.49 \textwidth]{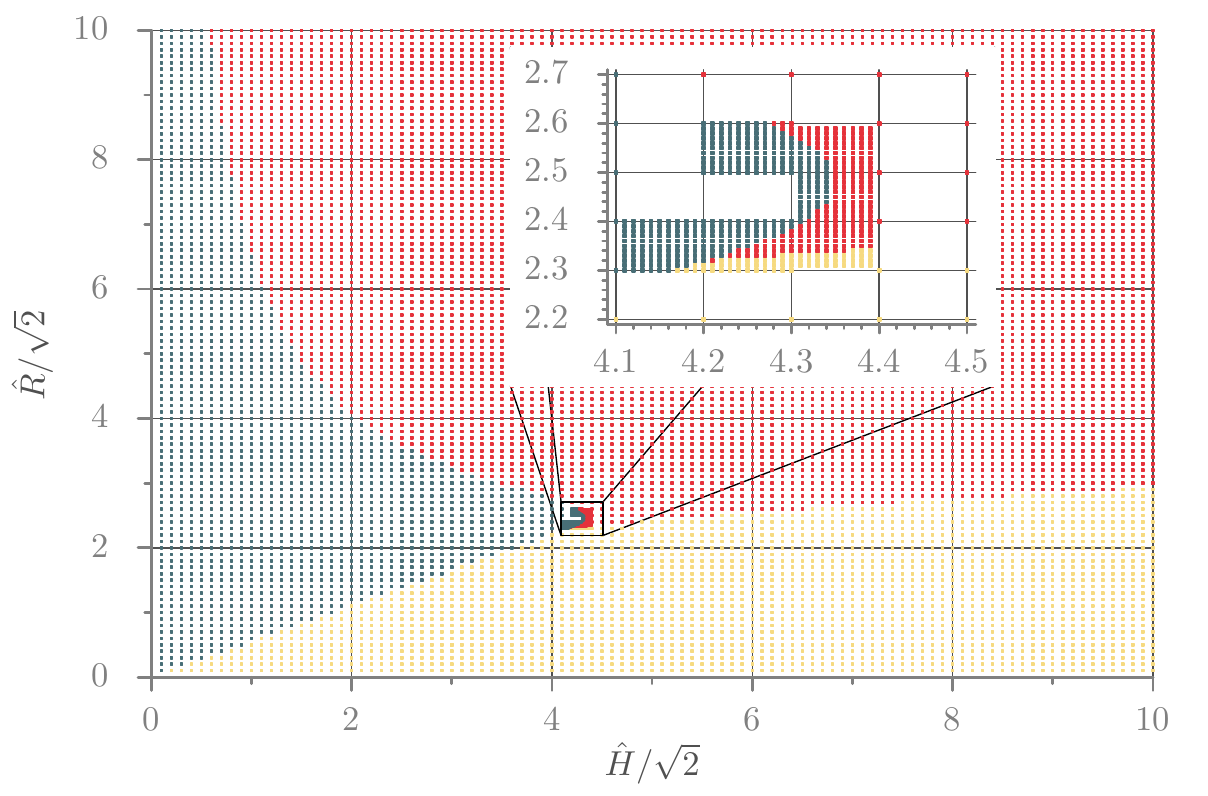}
    \includegraphics[page=1,width=0.49 \textwidth]{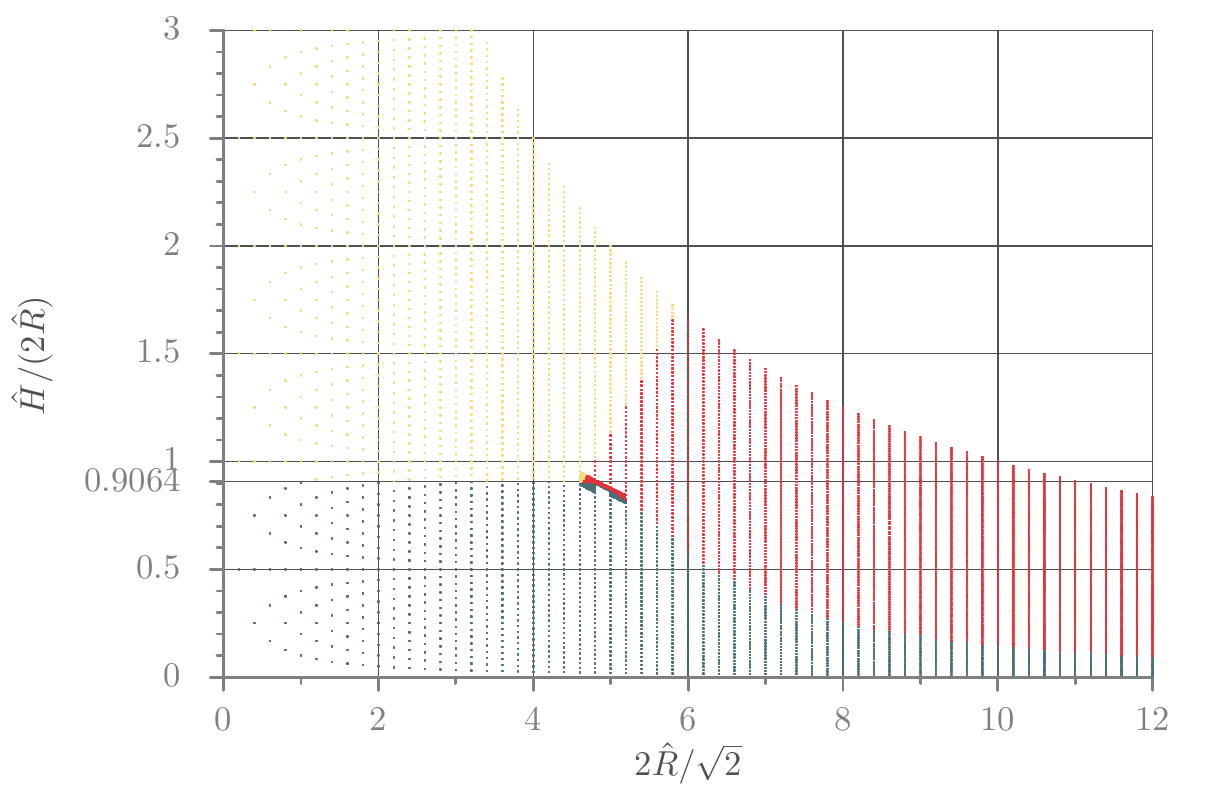}
	\caption{Results obatined by minimizing~\cref{eq:energy_f_nanodot} with the ansatz~\cref{eq:ritz_fourier} and by comparing the minimizing energy to the energy of the in-plane state \cref{eq:totalEnergyStrayNanoDot_final}. We sampled w.r.t.\ the radius and the height by using a grid of \(100\times100\) data points. Additionally, we sub-sampled the at a height of \(4.2\) and a radius of \(2.2\) to capture the character of the boundary separating the individual states with higher precision. Each of the samples is colored depending on its minimizing configuration type given in~\cref{fig:configurations}. Here, green refers to a homogeneous in-plane magnetization, red refers to a curling mode, and yellow refers to a homogeneous, out-of-plane magnetization.}
	\label{fig:Minimizer_of_modes_withData}
\end{figure}

To show the favourable minimizers in dependence of the nanodot's thickness and radius, we sampled the minimizer of \cref{eq:energy_f_nanodot} with our ansatz \cref{eq:ritz_fourier} and compared it with~\cref{eq:totalEnergyStrayNanoDot_final}. The results of this procedure were then assembled in a phase diagram taking into account the three different configurations given in~\cref{fig:configurations}, see~\cref{fig:Minimizer_of_modes_withData}. Here, the colors indicate the respective mode that provides the lowest energy minimum. The similarity of our results with~\cite[Fig.~2]{budaMicromagneticSimulationsMagnetisation2002} w.r.t.\ the border between the vortex configuration and the in-plane single-domain state is obvious. However,~\cite{budaMicromagneticSimulationsMagnetisation2002} takes the full three-dimensional problem into account.

Motivated by~\cite{rossMicromagneticBehaviorElectrodeposited2002}, we also show the sampled data in a recast format in the same figure. Comparing the latter representation with \cite[Fig. 6]{rossMicromagneticBehaviorElectrodeposited2002} indicates that in particular the border between the out-of-plane and the in-plane configuration as well as the border between the vortex and the in-plane configuration are nicely captured.

We note that the border between the out-of-plane configuration and the in-plane configuration can be determined exactly by comparing the corresponding energies~\cref{eq:energy_f_nanodot_out-of-plane3} and~\cref{eq:totalEnergyStrayNanoDot_final} such that
\begin{align}
	\frac{\widehat{\PI}^{\text{IP}}_\text{stray}}{\widehat{\PI}^{\text{OOP}}_\text{stray}} \stackrel{!}{=} 1 .
	\label{eq:energy_f_nanodot_border_IP_OOP}
\end{align}
This leads to the equation
\begin{align}
	3\pi\sqrt{\tau^2+1}\hypergeom{2}{1}\left(-\frac{1}{2}, \frac{3}{2};2;\frac{1}{\tau^2+1}\right)-2\pi\tau = 4 ,
	\label{eq:energy_f_nanodot_border_IP_OOP_final}
\end{align}
which can be solved numerically for \(\tau=\widehat{H}/(2\widehat{R})=1/\beta\), providing \(\tau=0.9064\,7615\,0066\,0646\). This result was also derived in \cite{aharoniSingledomainFerromagneticCylinder1989} by comparing the corresponding demagnetization factors of the two configurations. The value for $\tau$ can be also found explicitly or implicitly in, e.g., \cite{tandonComputationDemagnetizationTensor2004,rossMicromagneticBehaviorElectrodeposited2002,guslienkoReorientationalMagneticTransition}. Note that the value for $\tau$ can also be nicely seen in~\cref{fig:Minimizer_of_modes_withData}, where it separates the in-plane and the out-of-plane single-domain states.

For completeness, we mention that the border between the out-of-plane single-domain state and the vortex mode is quite different to the one reported by~\citet{rossMicromagneticBehaviorElectrodeposited2002}. This can be explained by the fact that in~\cite{rossMicromagneticBehaviorElectrodeposited2002} a three-dimensional micromagnetic simulation is carried out, which captures the out-of-plane flower state. In addition, the experimental data documented in~\cite{rossMicromagneticBehaviorElectrodeposited2002} has been obtained from arrays of \textit{almost} cylindrical nanodots that are densely packed. Since we take into account individual nanodots with \textit{perfectly} cylindrical shape, a quantitative comparison seems implausible. As a final note, we would like to mention the apparent similarities with the simulation results shown in \cite[Fig. 9a]{chung2010}.

\section{Summary and outlook}
We investigated the variational formulation and corresponding minimizing energies for typical magnetization states of thin cylindrical magnetic nanodots. For that, we considered both the exchange and the demagnetization energy and used Ritz's method in conjunction with a Fourier series expansion to calculate the energy minimizers. By comparing the minimizers of the vortex mode to the energies of in- and out-of-plane single-domain states, we derived a phase diagram of the nanodot and compared it to data obtained from two- and three-dimensional models from the literature. Our results allowed us to determine the critical radius at which the vortex mode becomes unfavorable with arbitrary precision. Current research is devoted to establishing a sophisticated, three-dimensional, micromagnetic, numerical simulation framework that will be benchmarked against the results that have been obtained in the present study.

\section*{Data availability statement}
The data and the scripts for creating the data are available via~\cite{darus-3103_2023}. The software Ikarus is available at~\cite{darus-3303_2023}.

\section*{Acknowledgements}
We gratefully acknowledge the support for this work
by Deutsche Forschungsgemeinschaft (DFG, German Research Foundation) under Germany's Excellence Strategy -- EXC 2075 -- 390740016.

\bibliographystyle{cas-model2-names}


\end{document}